\documentclass[%
superscriptaddress,
 amsmath,amssymb,
]{revtex4-2}

\usepackage[a4paper,top=25mm,bottom=25mm,left=30mm,right=25mm]{geometry}
\usepackage{graphicx}
\usepackage{bm}

\def\##1{\underline #1}
\def\=#1{\underline{\underline #1}}

\def\epst{{\boldsymbol \epsilon}}

\def\epst{{\boldsymbol \epsilon}}

\def\epst{{\boldsymbol \epsilon}}

\usepackage{hyperref}
\usepackage{cleveref}
\hypersetup{
    colorlinks=true,
    linkcolor=blue,
    citecolor=blue,
    urlcolor=blue,
    breaklinks=true,
}

\begin{document}
\preprint{APS/123-QED}

\title{Chirality-driven all-optical image differentiation}

\author{Stefanos Fr.\ Koufidis}
\email{steven.koufidis20@imperial.ac.uk}
\affiliation{Blackett Laboratory, Department of Physics, Imperial College of Science, 
Technology and Medicine, Prince Consort Road, London SW7 2AZ, UK}
\author{Zeki Hayran}
\email{z.hayran@imperial.ac.uk}
\thanks{Joint first co-authorship.}
\affiliation{Blackett Laboratory, Department of Physics, Imperial College of Science, 
Technology and Medicine, Prince Consort Road, London SW7 2AZ, UK}

\author{Francesco Monticone}
\affiliation{School of Electrical and Computer Engineering, Cornell University, Ithaca, New York 14853, USA}
\author{John B. Pendry}
\affiliation{Blackett Laboratory, Department of Physics, Imperial College of Science, Technology and Medicine, Prince Consort Road, London SW7 2AZ, UK}
\author{Martin W.\ McCall}
\affiliation{Blackett Laboratory, Department of Physics, Imperial College of Science, Technology and Medicine, Prince Consort Road, London SW7 2AZ, UK}

\date{\today}

\begin{abstract}
Optical analog computing enables powerful functionalities, including spatial differentiation, image processing, and ultrafast linear operations. Yet, most existing approaches rely on resonant or periodic structures, whose performance is strongly wavelength-dependent, imposing bandwidth limitations and demanding stringent fabrication tolerances. Here, to address some of these challenges, we introduce a highly tunable platform for optical processing, composed of two cascaded uniform slabs exhibiting both circular and linear birefringence, whose response exhibits features relevant to optical processing without relying on resonances. Specifically, using a coupled-wave theory framework we show that sharp reflection minima, referred to as spectral holes, emerge from destructive interference between counter-propagating circularly polarized waves in uniform birefringent slabs, and can be engineered solely through parameter tuning without requiring any spatial periodicity. Unlike traditional Bragg scattering, this mechanism operates without a resonance condition and enables a comparatively broader spectral response through material parameter tuning in spatially uniform media. When operated in the negative refraction regime enabled by giant chirality, the proposed system acts as a polarization-selective Laplacian-like operator, whose functionality is evidenced by an edge-detection proof of concept. The required material parameters align closely with recent experimental demonstrations of giant, tunable chirality via meta-optics, presenting a promising pathway towards compact and reconfigurable platforms for all-optical pattern recognition and image restoration.
\end{abstract}

\maketitle

\section{Introduction} 
\label{Sec:Introduction}
Optical analog computing harnesses the wave nature of light to execute mathematical operations via passive, parallel physical processes, offering significant reductions in energy consumption and latency compared to conventional electronic computation, despite similar underlying signal speeds \cite{McMahon2023}. Early demonstrations showed that engineered meta-media could realize spatial operations such as edge detection and convolution by encoding mathematical kernels into the local amplitude and phase responses of planar nanostructures \cite{Silva2014,Xu2022}, with more recent implementations spanning from dispersion-engineered metasurfaces \cite{Cotrufo2023} to nonlinear thin films \cite{Cotrufo2025}. These passive, planar platforms carry out complex computations directly on incident wavefronts, highlighting their potential for real-time image processing, neuromorphic photonics, and other applications that exploit inherent spatial parallelism \cite{Hu2024}. 

Programmable and reconfigurable optical platforms have also recently been developed that can implement arbitrary linear—and even nonlinear—transformations, such as matrix inversion and root finding, thereby broadening the computational repertoire of wave-based processors \cite{Nikkhah2023,Tzarouchis2025}. Temporal metamaterials, in turn, provide a fundamentally distinct platform for analog computing, drawing on nonlocal effects \cite{Rizza2022} or spin-controlled dynamics \cite{Rizza2023} to achieve first-order waveform differentiation.

Complementary to these developments, analog optical operations have also been demonstrated using a range of other photonic mechanisms, often relying on distributed-feedback effects to manipulate the phase response of propagating light. Techniques based on phase-shifted gratings have enabled first- and higher-order differentiation and integration by embedding discrete phase discontinuities into otherwise homogeneous media \cite{Berger2007,Kulishov2007,Ngo2007}. Such methods have been successfully adapted to THz-compatible devices using silicon-on-insulator directional couplers \cite{Huang2015} and reconfigurable interferometric signal processors capable of performing versatile linear operations on-chip \cite{Babashah2019}. Related approaches involving phase-shifted Bragg gratings, plasmonic waveguides, or spatially dispersive metasurfaces composed of anisotropic unit cells \cite{Qiu2025} have enabled beam-profile differentiation by selectively reflecting angular field components, effectively implementing spatial operations in the Fourier domain \cite{Doskolovich2014,Bykov2014,Zhu2017, Tanriover2023}. Nonetheless, the reliance of these systems on resonant or periodic structures restricts their operation to narrow spectral bands and precise incidence angles, limiting their potential for general-purpose broadband computing. These constraints also impose demanding fabrication tolerances, hindering scalability and reconfigurability.

In this communication we propose a parameter-tunable, non-periodic mechanism for achieving broadband Laplacian-type differentiation in modulus, by cascading two uniform slabs of chiral-birefringent media. Each slab combines circular birefringence, characterized by a magnetoelectric chirality parameter $\alpha$, with linear dielectric anisotropy. By introducing a slight difference between the average refractive indices of the slabs, we derive explicit conditions under which destructive interference generates a sharp reflection dip—a ``spectral hole''—in the chirality domain. Coupled-wave analysis reveals that at this spectral hole the magnitude of the reflection coefficient exhibits a parabolic dependence on the transverse wavenumbers, $k_x$ and $k_y$, thereby yielding a transfer function proportional to $k_x^2 + k_y^2$, the hallmark of Laplacian-like differentiation, over a broad spectral range. 

The required parameters align closely with recent advances in metamaterials \cite{Zhang2009} and metasurfaces \cite{Wu2023} exhibiting giant and controllable chirality \cite{Liu2021}. Consequently, our approach yields compact, reconfigurable photonic differentiators that are not subject to the narrowband spectral response imposed by the resonant wavelength dependence of periodic structures governed by the Bragg condition. As with all related methods discussed above, our approach is naturally suited for operation with spatially coherent light \cite{Swartz2024}.

The manuscript is organized as follows: After revisiting the eigenmodes of circularly and linearly birefringent media, detailed in App.\ \ref{App:Cascaded Circularly and Linearly Birefringent Media}, Sec.\ \ref{Sec:Chiral Spectral Holes} delineates the conditions necessary for generating high-quality spectral holes in the chirality domain, as opposed to the traditional wavelength domain. Subsequently, Sec.\ \ref{Sec:Broadband Spatial Laplacian-like Differentiation} evaluates the functionality and performance of the proposed polarization-selective spatial differentiator, demonstrating the excellent parabolic fit of the transfer function and illustrating its utility in edge detection applications. Furthermore, by accounting for realistic material dispersion, we directly compare the proposed scheme with established Bragg-based devices, demonstrating its superior performance in terms of angular range. Section \ref{Sec:Meta-Media Implementation} then discusses practical implementations in current meta-media that satisfy the broadband operational criteria, followed by Sec.\ \ref{Sec:Conclusions} which summarizes our key findings.

\section{Chiral Spectral Holes}
\label{Sec:Chiral Spectral Holes}

\begin{figure}[h]
\centering
\includegraphics[width=0.5\linewidth]{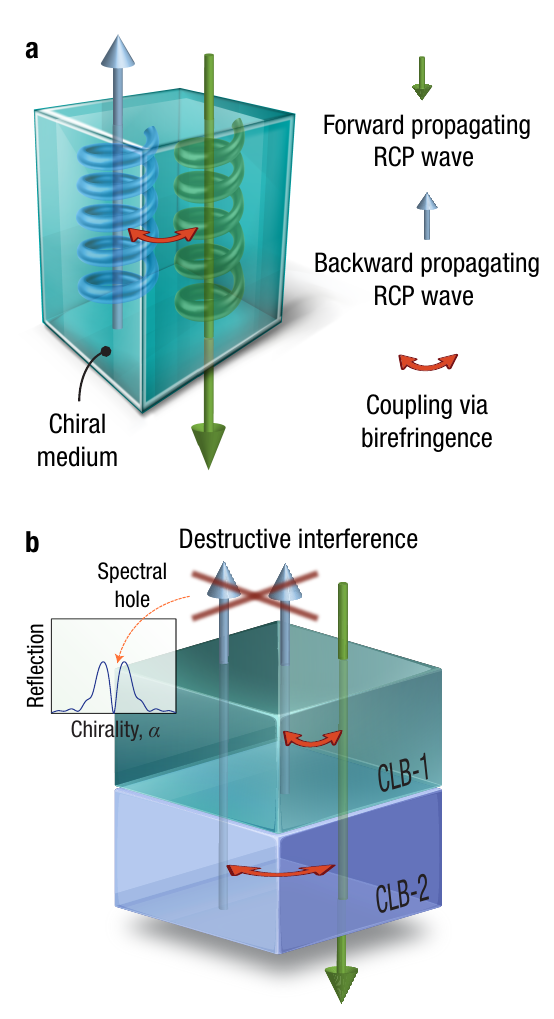}
\caption{\textbf{Chirality-induced reflection effects in birefringent media.}  
(a) Interaction between forward- and backward-propagating circularly polarized waves in a chiral medium, mediated by linear birefringence, gives rise to the notion of a ``grating-less grating'' reported in Ref.\ \cite{Koufidis2023a}.
(b) The structure under discussion consists of two \emph{uniform} slabs of equal thickness, labeled CLB$_1$ and CLB$_2$, each exhibiting both circular birefringence and transverse dielectric anisotropy. Under corresponding preconditions, a slight mismatch in their average refractive indices leads to destructive interference between counter-propagating modes. This interference results in a reflection minimum that appears in the chirality domain, governed entirely by material parameter tuning rather than spatial periodicity. Since this response does not originate from a conventional resonance, the effect is not constrained by a narrow resonance bandwidth, but only by material dispersion.}
\label{Fig_Schematic}
\end{figure}

Traditionally, the concept of all-optical spatial differentiation has been primarily associated with the formation of spectral holes in the optical spectrum of cascaded, spatially modulated Bragg gratings~\cite{Doskolovich2014, Bykov2014}. At the core of Bragg gratings lies the Bragg condition, whereby incident lightwaves whose wavelength matches the pitch of the grating are strongly backscattered, as successive reflections coherently superimpose. However, Ref.\ \cite{Koufidis2023a} introduced a distinct yet physically related mechanism that arises in \emph{homogeneous} media exhibiting both circular and linear birefringence (CLB), thereby lifting the requirement for wavelength matching---see Fig. \ref{Fig_Schematic}(a). Owing to its dependence solely on material parameters, the recently identified Bragg-like effect is intrinsically broadband, limited only by material dispersion.

A schematic depiction of a configuration capable of supporting chiral spectral holes is shown in Fig.~\ref{Fig_Schematic}(b): it comprises two cascaded, homogeneous slabs---CLB$_1$ and CLB$_2$---each exhibiting both circular and linear birefringence. A circularly polarized wave of a given handedness, say right, is normally incident onto the system and is partially coupled into a backward-propagating wave through birefringence-induced mixing. As demonstrated in the analysis below, a slight mismatch between the average refractive indices of the two slabs ensures destructive interference between the counter-propagating waves. Such an interference produces a pronounced reflection minimum---i.e., a spectral hole---that originates from the intrinsic material parameters and the slab length, rather than from any wavelength-scale periodic modulation. 

To begin with, let us consider a medium characterized by relative permittivity $\epsilon_{\mathrm{r}}$, permeability $\mu_{\mathrm{r}}$, and chirality parameter $\alpha$. The latter quantifies magneto-electric coupling, inducing circular birefringence by shifting the refractive indices of right- and left-circular polarizations (RCP and LCP, respectively) by $\mp\alpha$. Physically, the chirality parameter $\alpha$ corresponds to the propagation distance, in wavelengths, over which the polarization vector completes a full $2\pi$ rotation. As discussed in App.~\ref{App:Cascaded Circularly and Linearly Birefringent Media}, circular birefringence induced by magneto-electric coupling due to $\alpha\neq0$ results in distinct refractive indices for right- and left-circularly polarized light, namely $n_{\mathrm{R}} = n - \alpha$ and $n_{\mathrm{L}} = n + \alpha$, where $n = (\epsilon_{\mathrm{r}} \mu_{\mathrm{r}})^{1/2}$ is the background refractive index.

When circular birefringence is combined with linear birefringence—arising from an anisotropic permittivity, $\boldsymbol{\epsilon}_\bot = \operatorname{diag}(\epsilon_1, \epsilon_2)$—the resulting medium supports coupled-wave dynamics analogous to those found in uniform Bragg gratings (see App.~\ref{App:Cascaded Circularly and Linearly Birefringent Media}). By contrast to conventional gratings, however, resonance now depends solely on the intrinsic parameters of the medium and the slab length, rather than on subwavelength spatial periodicity or lithographic patterning. Specifically, for a circularly and linearly birefringent medium, phase-matching leads to a ``resonant'' condition $\alpha = \pm \bar{n}$. Here $\bar{n} = (\bar{\epsilon}_{\mathrm{r}} \mu_{\mathrm{r}})^{1/2}$ denotes the average refractive index and $\bar{\epsilon}_{\mathrm{r}} = (\epsilon_1 + \epsilon_2)/2$ is the mean permittivity; unless stated otherwise, $\mu_{\mathrm{r}}=1$.

Assuming $\bar{n}$ to be dispersionless, the detuning $\delta_{R,L} = 2k_0\left(\bar{n} \mp \alpha\right)$, where $k_0$ is the free-space wavenumber, becomes wavelength-independent and determined solely by the medium parameters (a discussion of dispersion and realistic implementation using meta-media is deferred to Sec.\ \ref{Sec:Meta-Media Implementation}). Furthermore, each resonant condition (centered at either $\alpha = +\bar{n}$ or $\alpha = -\bar{n}$) is associated with a distinct handedness. Indeed, the condition $\delta_{\mathrm{R}} = 0$ (respectively, $\delta_L = 0$) yields a polarization-selective reflective response centered at $\alpha = \bar{n}$ (respectively, $\alpha = -\bar{n}$) under RCP (respectively, LCP) light excitation. In the vicinity of this regime \emph{all} wavelengths are reflected, insofar as $\left|\alpha(\lambda) - \bar{n}(\lambda)\right| \approx 0$, motivating the concept of a spatially uniform Bragg reflector \cite{Koufidis2023a}. Yet, in the absence of spatial periodicity, one may ask: how can a spectral hole arise?

It turns out that the answer lies not in the wavelength domain but in the domain of chirality. To show this, let us consider two cascaded slabs of equal length, each comprising homogeneous media exhibiting both circular and linear birefringence, as depicted in Fig. \ref{Fig_Schematic}(b). The first slab (CLB$_1$), of length $L/2$, has principal permittivities $(\epsilon_1, \epsilon_2)$ and average refractive index $\bar{n}_1 = \left[(\epsilon_1 + \epsilon_2)/2\right]^{1/2}$. The second slab (CLB$_2$), also of length $L/2$, has principal permittivities $(\epsilon_2, \epsilon_3)$ and average refractive index $\bar{n}_2 = \left[(\epsilon_2 + \epsilon_3)/2\right]^{1/2}$; importantly $\epsilon_1 \neq \epsilon_3$.

For, say, RCP excitation, each slab is characterized by a distinct set of coupled-wave theory parameters: detuning $\delta_{\mathrm{R}}^{(j)}$ and coupling coefficient $\kappa^{(j)}$; $j=\{1,2\}$. In particular, CLB$_1$ has $\delta_{\mathrm{R}}^{(1)} = 2k_0\left(\bar{n}_1 - \alpha\right)$ and $\kappa^{(1)} = \left({k_0}/{2}\right)\left|\epsilon_1^{1/2} - \epsilon_2^{1/2}\right|$, whereas CLB$_2$ has $\delta_{\mathrm{R}}^{(2)} = 2k_0\left(\bar{n}_2 - \alpha\right)$ and $\kappa^{(2)} = \left({k_0}/{2}\right)\left|\epsilon_2^{1/2} - \epsilon_3^{1/2}\right|$. Accordingly one can construct a transfer matrix $\mathbf{T}_{\mathrm{R}}^{(j)}$ for each slab. These matrices relate the electric field amplitudes at $z_{j}$ to those at $z_{j-1}$ via $\mathbf{A}_{\mathrm{R}}(z_{j}) = \mathbf{T}_{\mathrm{R}}^{(j)} \cdot \mathbf{A}_{\mathrm{R}}(z_{j-1})$, where $\mathbf{A}_{\mathrm{R}}(z_{j}) = \left(A_{\mathrm{R}}^+(z_{j}), A_{\mathrm{R}}^-(z_{j})\right)^\intercal$, with $^\intercal$ denoting transpose and ``$\pm$'' indicating the direction of phase propagation. The transfer matrix of the cascaded structure is 
\begin{equation}\label{Total_T_Matrix}
    \mathbf{T}_{\mathrm{R}} =\mathbf{T}_{\mathrm{R}}^{(2)} \mathbf{T}_{\mathrm{R}}^{(1)}= \begin{pmatrix} T_{11} & T_{12} \\ 
    T_{21} & T_{22} \end{pmatrix}\,.
\end{equation}
Its components, whose analytic expressions are provided in App.\ \ref{App:Cascaded Circularly and Linearly Birefringent Media}, depend on the coupled-wave parameters: 
      \begin{align}
    p_{\mathrm{R}}^{\pm,(j)} &= \cosh\left(\Delta_{\mathrm{R}}^{(j)} L_{j}^{-}\right) \pm i \frac{\delta_{\mathrm{R}}^{(j)}}{2\Delta_{\mathrm{R}}^{(j)}} \sinh\left(\Delta_{\mathrm{R}}^{(j)} L_{j}^{-}\right)\,, \label{CWT_P}
    \\
    q_{\mathrm{R}}^{\pm,(j)} &= \pm i\frac{\kappa^{(j)}}{\Delta_{\mathrm{R}}^{(j)}} \sinh\left(\Delta_{\mathrm{R}}^{(j)} L_{j}^{-}\right)\,, \label{CWT_Q}
\end{align}  
with $\Delta_{\mathrm{R}}^{(j)} = \left[\left(\kappa^{(j)}\right)^2 - \left(\delta_{\mathrm{R}}^{(j)}/2\right)^2\right]^{1/2}$ and $L_j^{\pm}=z_{j} \pm z_{j-1}$ highlighting the dependence of $\mathbf{T}_{\mathrm{R}}$ on $z_{j-1}$ \cite{McCall2000}.

For incident RCP light, imposing the boundary condition $A_{\mathrm{R}}^-\left(z=L\right) = 0$ yields the right-to-right reflection and transmission coefficients: $r_{\mathrm{RR}} = -T_{21}/T_{22}$ and $t_{\mathrm{RR}} = 1/T_{22}$, respectively. In order to isolate the reflection effects arising solely from the simultaneous presence of the two forms of birefringence, we hereafter assume that the composite finite slab is embedded in a surrounding medium of refractive index $\bar{n}$, such that approximate index matching is achieved. Variations in transverse birefringence or in the effective slab thickness that could induce Fabry–P\'erot phase shifts are presently neglected.

For a spectral hole to arise at a specific wavelength $\lambda_{0}^{\rm s.h.}$, two \emph{counter}-propagating eigenmodes must interfere destructively, i.e.,  $r_{\mathrm{RR}}\left(\lambda_{0}^{\rm s.h.}\right) = 0$. This implies that $T_{21} = 0$, which upon substitution of the expressions for $p_{\mathrm{R}}^{\pm,(j)}$ and $q_{\mathrm{R}}^{\pm,(j)}$ into the formula for $T_{21}$ given in App.\ \ref{App:Cascaded Circularly and Linearly Birefringent Media} yields
\begin{align}\label{Gen_Spect_Hole}
     &e^{i\left(\delta_{\mathrm{R}}^{(2)}-\delta_{\mathrm{R}}^{(1)}\right) L/2} =-\frac{\kappa^{(1)} \tanh{\left( \Delta_{\mathrm{R}}^{(1)} L/2 \right)}}{\kappa^{(2)} \tanh{\left( \Delta_{\mathrm{R}}^{(2)} L/2 \right)}}\frac{2\Delta_{\mathrm{R}}^{(2)} - i \delta_{\mathrm{R}}^{(2)} \tanh{\left( \Delta_{\mathrm{R}}^{(2)} L/2 \right)}}{2\Delta_{\mathrm{R}}^{(1)} + i \delta_{\mathrm{R}}^{(1)} \tanh{\left( \Delta_{\mathrm{R}}^{(1)} L/2 \right)}}\,.
\end{align}

It follows from Eq. \eqref{Gen_Spect_Hole} that the left-hand side simplifies to $e^{ik_0(\bar{n}_1 - \bar{n}_2)L}$. Whence, for Eq. \eqref{Gen_Spect_Hole} to admit a solution for some finite value of $L$, the right-hand side must equal $-1$. This condition is met only if \emph{both} media are tuned to \emph{precisely} the same chirality parameter---a trivial scenario corresponding to unitary spectral hole transmission. To access a non-trivial regime, we perturb around this degenerate case while still retaining unit transmittance. Concretely, we consider a nearly coincident configuration in which the refractive indices of CLB$_1$ and CLB$_2$ differ \emph{slightly}. Specifically, we assume that $\left|\epsilon_2^{1/2} - \epsilon_1^{1/2}\right|$ and $\left|\epsilon_3^{1/2} - \epsilon_2^{1/2}\right|$ are sufficiently small such that each slab is resonant at nearly, but not exactly, the same value of $\alpha$. Under this near-resonant condition, Eq. \eqref{Gen_Spect_Hole} reduces approximately to $e^{ik_0\Delta n_{21}L_c} \approx -1$, leading to the characteristic structure length for destructive interference
\begin{equation}\label{Chir_Spec_Hol_Length}
L_c \approx \frac{\lambda_0}{2\Delta n_{21}}\,,
\end{equation}
where $\Delta n_{21} = \bar{n}_2 - \bar{n}_1$ quantifies the index mismatch.

\begin{figure*}[!t]
\centering
\includegraphics[width=\linewidth]{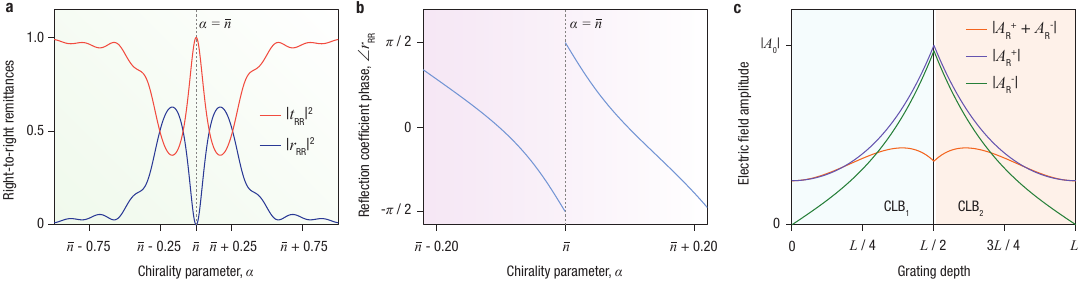}
\caption{\textbf{Chirality-domain spectral hole formation in cascaded circularly and linearly birefringent media.}  
Optical response of the structure shown in Fig.~\ref{Fig_Schematic}(b), when embedded in an index-matched surrounding medium of refractive index $\bar{n}$. (a) Right-to-right intensity reflectance and transmittance (remittances) as functions of the chirality parameter $\alpha$. The generalized interference condition of Eq.~\eqref{Gen_Spect_Hole}, combined with the characteristic slab length approximated by Eq.~\eqref{Chir_Spec_Hol_Length}, leads to complete destructive interference at $\alpha = \bar{n} = (\bar{n}_1 + \bar{n}_2)/2$, thereby generating a spectral hole in the chirality domain. By contrast to spatially periodic gratings, this mechanism requires no spatial modulation, and the resulting reflection minimum does not originate from a conventional resonance, avoiding the sharp spectral constraints typically associated with the Bragg condition of wavelength-resonant systems. The structure also exhibits polarization selectivity analogous to the circular Bragg phenomenon in structurally chiral media. (b) Phase response under the same conditions as in panel (a), featuring a sharp phase transition at the spectral hole, thus signifying the ``resonant'' character (in the chirality domain) of the reflection zero. (c) Spatial profiles of the forward- and backward-propagating electric field amplitudes within the structure at $\alpha = +\bar{n}$. The fields are strongly localized near the interface between the slabs, with the backward-propagating component decaying downstream, consistent with the vanishing reflectance and near-unity transmittance observed in panel (a), without any attempt on impedance matching. Simulation parameters follow Eq.~\eqref{Permittivities Triples}, with $\epsilon_1 = 4$ and $\delta_\epsilon = 2$ used as representative values.}

\label{Fig_Opt_Response}
\end{figure*}

This relation enables the design of a uniform structure supporting chiral spectral holes through an appropriate selection of the triplet $(\epsilon_1, \epsilon_2, \epsilon_3)$. For ordering such that $\epsilon_1<\epsilon_2<\epsilon_3$ we may impose as a design strategy that $\kappa^{(1)} = \kappa^{(2)}$, whence we obtain
\begin{equation*}
2\epsilon_2^{1/2} = \epsilon_1^{1/2} + \epsilon_3^{1/2}.
\end{equation*}
Given a base permittivity $\epsilon_1$ and a perturbation $\delta_\epsilon > 0$, the triplet can then be cast as
\begin{equation}\label{Permittivities Triples}
\epsilon_1\,, \quad \epsilon_3 = \epsilon_1 + \delta_\epsilon\,, \quad \text{and} \quad \epsilon_2 = \left[ \frac{1}{2} \left( \epsilon_1^{1/2} + \epsilon_3^{1/2} \right) \right]^2\,.
\end{equation}
By systematically varying $\delta_\epsilon$, one can control the separation between $\epsilon_1$ and $\epsilon_3$, thereby enabling precise numerical tuning while satisfying the constraint of Eq.\ \eqref{Chir_Spec_Hol_Length}. The intensity reflectance and transmittance spectra of such an optimized structure are shown in Fig.\ \ref{Fig_Opt_Response}(a), whereas Fig.\ \ref{Fig_Opt_Response}(b) displays the corresponding phase response. As opposed to conventional optical spectra, the abscissa here represents variations in chirality rather than wavelength.

This distinction is further evidenced in Fig.\ \ref{Fig_Opt_Response}(c), which plots the electric field amplitude distribution within the structure as a function of the grating depth. Rather than the forward-traveling wave being strongly coupled into the backward-traveling wave, the latter attains its peak at the interface of the two cascaded gratings and subsequently decays as it propagates through the remainder of the structure. By contrast to phase-shifted Bragg gratings, where this behavior is attributed to the wavelength falling within a spectral hole, the localization observed here stems from the design strategy itself, i.e., the choice of parameters satisfying Eqs.\ \eqref{Chir_Spec_Hol_Length} and \eqref{Permittivities Triples}, upon which the chiral tuning ultimately hinges.

\section{Broadband Spatial Laplacian-like Differentiation}
\label{Sec:Broadband Spatial Laplacian-like Differentiation}

An all-optical Laplacian-like differentiator imparts a spatially uniform second-order response across the angular spectrum of an incident beam, such that the reflected (or transmitted) field closely approximates the Laplacian of the input wavefield profile in modulus. Realizing this functionality requires engineering the structure’s transfer function---that is to say, the reflection coefficient $r_{\mathrm{RR}}$---to exhibit a quadratic dependence on the transverse wavenumbers $k_x$ and $k_y$. Capitalizing on the principles established in Sec.~\ref{Sec:Chiral Spectral Holes}, we demonstrate in this section that Laplacian-like differentiation can be achieved via the setup of Fig.\ \ref{Fig_Schematic}(b) by exploiting spectral holes arising in the chirality domain.

\begin{figure}[h]
\centering
\includegraphics[width=0.5\linewidth]{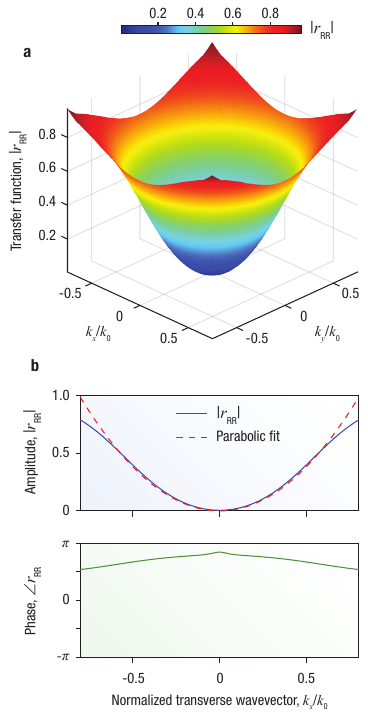}
\caption{
{\bf Transfer function features of the chirality-driven, all-optical Laplacian differentiator in modulus.}
(a) Amplitude of the reflection coefficient, $|r_{\mathrm{RR}}|$, for an incident RCP plane wave, plotted as a function of the normalized transverse wavenumbers $k_x/k_0$ and $k_y/k_0$. The resulting cone-like profile exhibits a sharp minimum at $k_x = k_y = 0$, consistent with the destructive interference condition $r_{\mathrm{RR}}(\lambda, \alpha) = 0$ derived in Sec.~\ref{Sec:Chiral Spectral Holes}. The amplitude follows a symmetric quadratic dependence, signature of Laplacian-type behaviour. 
(b) Amplitude (top panel) and phase (bottom panel) of the reflection coefficient as functions of $k_x/k_0$. The amplitude curve is fitted with a parabola (dashed red line), whose curvature determines the proportionality constant $C$ in Eq.~\eqref{Refl_Field_Laplacian}. The phase response varies smoothly across the reflection minimum without discontinuities, indicating a high-fidelity implementation of a Laplacian operator in modulus acting on the angular spectrum of the incident field.
}
\label{Fig_Transfer_Function}
\end{figure}

In principle Laplacian differentiation emerges from the steep parabolic dependence of the reflection coefficient in the vicinity of its zero. The condition of destructive interference, $r_{\mathrm{RR}}(\lambda,\alpha)=0$, defines a spectrally sharp reflection minimum wherein both the magnitude $|r_{\mathrm{RR}}|$ and the phase $\angle\bigl(r_{\mathrm{RR}}\bigr)$ vary rapidly yet in a symmetric fashion. For an incident plane wave, Fig.\ \ref{Fig_Transfer_Function}(a) depicts the three-dimensional surface of the reflection coefficient amplitude as a function of the normalized transverse wavenumbers $k_x/k_0$ and $k_y/k_0$. The resulting parabolic surface profile, proportional to $k_x^2 + k_y^2$, clearly exhibits the hallmark characteristic of a Laplacian operator in modulus across the transverse spatial frequency domain. Focusing on the $k_x$-axis, the top panel of Fig.\ \ref{Fig_Transfer_Function}(b) shows that the amplitude of the reflection coefficient is approximately symmetric and quadratic in $k_x$, overlaid with a parabolic fit, where the remarkable agreement even for high values of $k_x$ underscores both the quality of the engineered spectral hole and the overall performance of the Laplacian-like operator. The corresponding phase response, shown in the bottom panel of Fig.\ \ref{Fig_Transfer_Function}(b), reveals that near the zero-reflection point, the phase evolves continuously and without discontinuity. Crucially, while the slab length sets a design wavelength, the formation of the spectral hole is governed by material parameter tuning rather than a conventional resonance condition. In this context \emph{high}-quality pertains to the \emph{exact} fulfillment of the condition $r_{\mathrm{RR}}\bigl(\lambda,\alpha = \pm\bar{n}\bigr)=0$ across the relevant (i.e., operational) spectral range.

Mathematically, such a Laplacian-like optical response can be cast in terms of an effective operator acting on the input field. Indeed, let us consider a paraxial beam with transverse profile $E_i(x,y)$ incident on the structure. The angular spectrum of the reflected field is
\begin{equation*}
    E_{\mathrm{r}}(x,y) = \iint_{-\infty}^{\infty} r_{\mathrm{RR}}(k_x,k_y)\,\tilde{E}_i(k_x,k_y)\,e^{i(k_x x + k_y y)}\,\mathrm{d}k_x\,\mathrm{d}k_y,
\end{equation*}
where $\tilde{E}_i(k_x,k_y)$ is the 2D Fourier transform of $E_i(x,y)$. If the transfer function satisfies $r_{\mathrm{RR}}(k_x,k_y)\propto -\bigl(k_x^2 + k_y^2\bigr)$, then the reflected field approximates the modulus of the Laplacian of the input profile,
\begin{equation}\label{Refl_Field_Laplacian}
    E_{\mathrm{r}}(x,y)\approx -\,C\,\nabla^2 E_i(x,y)\,,
\end{equation}
with $C$ being a proportionality constant determined by the curvature of $r_{\mathrm{RR}}$ in the vicinity of the spectral hole. 

Since tuning is achieved through parameter matching rather than conventional Bragg resonance, the quadratic approximation of $r_{\mathrm{RR}}(k_x, k_y)$ near the reflection minimum does not rely on a narrowband resonant condition. Instead, it remains valid over a finite spectral region around the design wavelength determined by the slab length. Although the range is not inherently wavelength-independent, it is not constrained by the sharp spectral selectivity typical of resonant structures. If the dispersion of $\alpha$ and $\bar{n}$ is closely matched, the detuning remains small across a finite bandwidth, preserving the Laplacian-like response within practical limits. The curvature of the phase response near the reflection zero can be adjusted through the relative permittivity contrast, specified by the design parameter $\delta_\epsilon$ in Eq.~\eqref{Permittivities Triples}, offering additional control over the angular response of the differentiator.

Furthermore, a salient feature of our mechanism lies in its intrinsic polarization selectivity. Indeed, the transfer function $r_{\mathrm{RR}}(k_x,k_y)$ is inherently handedness-dependent: for a given sense of optical rotation, only one circular polarization state (RCP or LCP) experiences the chiral stopband, while the opposite handedness propagates unaffected, subject only to absorption. This selectivity arises from the symmetry breaking induced by the chirality parameter $\alpha$, which couples asymmetrically to the two orthogonal states. Hence, the structure performs Laplacian differentiation in modulus exclusively on the selected polarization component, enabling polarization-resolved spatial processing. If a linearly polarized beam—comprising equal amounts of RCP and LCP light—is incident, only the component aligned with the chosen chirality sign is differentiated in the corresponding channel.

\begin{figure*}[!t]
\centering
\includegraphics[width=\linewidth]{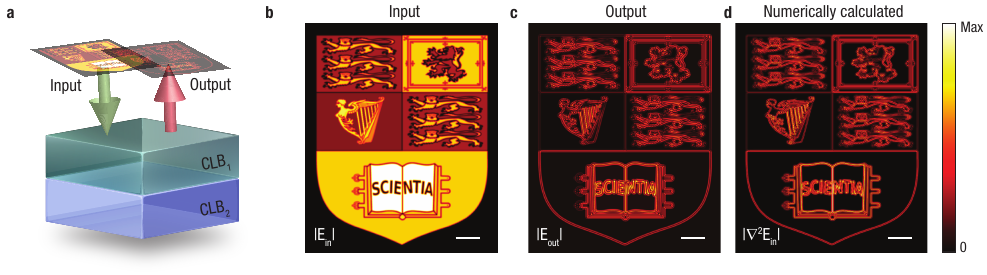}
\caption{
{\bf Chirality-driven, all-optical Laplacian operator in modulus enabling broadband, high-fidelity edge detection.} (a) Schematic of the proposed Laplacian-like differentiator: for the configuration displayed in Fig.\ \ref{Fig_Schematic}(b), when surrounded by an isotropic medium of refractive index $\bar{n}$, the reflected signal corresponds to the modulus of the Laplacian of the electric field profile of the incident lightwave.
(b) Normalized intensity profile of the input beam, $|E_\mathrm{in}|$, used to excite the structure.  
(c) Simulated amplitude of the reflected field, $|E_\mathrm{out}|$, based on the transfer function illustrated in Fig.~\ref{Fig_Transfer_Function}. The output accurately reproduces the high-contrast features expected from Laplacian-based edge detection.  
(d) Numerically calculated magnitude of the Laplacian of the input profile, $|\nabla^2 E_\mathrm{in}|$, shown for comparison. The proposed scheme accurately reconstructs the edges and fine details of the Imperial College crest, in excellent agreement with those obtained from the analytical Laplacian [cf.\ panels (c) and (d)]. The scale bars in panels (b)-(d) correspond to $20\times 2\pi/k_0$.
}
\label{Fig_Edge_Detection}
\end{figure*}

To illustrate the practical capabilities of the proposed differentiating scheme, Fig.\ \ref{Fig_Edge_Detection} presents simulations using the Imperial College crest as the input field profile [see Fig. \ref{Fig_Edge_Detection}(b)]. The structure used in Fig.~\ref{Fig_Edge_Detection}(a) corresponds to the configuration in Fig.~\ref{Fig_Schematic}(b), designed to produce a high-quality reflection zero at the target chirality parameter $\alpha$. Upon illumination, the reflected field reconstructs the Laplacian of the input image in modulus, highlighting edges and fine features with high fidelity. The output amplitude closely reproduces the numerically simulated one across the entire field of view [cf.\ panels (c) and (d) in Fig.\ \ref{Fig_Edge_Detection}], confirming the precision of chirality-domain tuning. Further simulations have confirmed that such a behaviour remains stable under variations in slab thickness and the chirality parameter. The output edge-enhanced images demonstrate that high-contrast, polarization-selective spatial processing can be realized with the proposed configuration, providing a scalable alternative to conventional grating- or metasurface-based edge detectors.

\begin{figure}[h]
\centering
\includegraphics[width=0.5\linewidth]{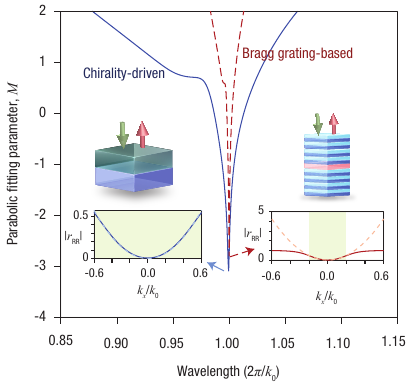}
\caption{
{\bf Comparison of operational bandwidths between the proposed chirality-driven and traditional Bragg grating-based differentiators.}
Parabolic fitting parameter $M$ is plotted as a function of normalized wavelength for both the chirality-driven (blue solid line) and Bragg grating-based (red dashed line) spatial differentiators. Insets display the reflection coefficient magnitude as a function of normalized transverse wavenumber ($k_x/k_0$) for each case at the operational wavelength $2\pi / k_0$, with solid curves representing the simulated data and dashed curves indicating the corresponding parabolic fits. The chirality-driven device exhibits a well-matched quadratic response over a broad angular range, $k_x/k_0 \in [-0.6, 0.6]$, whereas the 17-layer Bragg grating design of Ref.~\cite{Doskolovich2014} maintains parabolic behaviour only within the narrower interval $k_x/k_0 \in [-0.2, 0.2]$. As elaborated in App.\ \ref{App:Parabolic‑Fitting Parameter M}, lower values of $M$ correspond to a better parabolic approximation of the transfer function. The dispersive material properties for the chirality-driven differentiator are given in App.\ \ref{App:Lorentzian Material Dispersion}.
}
\label{Fig_Bandwidth}
\end{figure}

Finally, in Fig.~\ref{Fig_Bandwidth}, we benchmark the performance of the chirality-driven spatial differentiator against that of a conventional Bragg grating device, based on the 17-layer configuration proposed in Ref.~\cite{Doskolovich2014}. Since dispersion is inherently linked to negative refraction, we incorporate a physically consistent comparison by modeling the chirality-based structure using Lorentz-type dispersion for both the permittivity and chirality parameters [see Eqs.\ \eqref{L_Perm_x}, \eqref{L_Perm_y}, and \eqref{L_Chir} in App.~\ref{App:Lorentzian Material Dispersion}]. When negative refraction is present, dispersion is inevitable, since any stable system must possess a positive electromagnetic energy density. This requirement can be established by equating the energy flow, determined from the group velocity $v_g = \left|{{\rm d}k}/{{\rm d}\omega}\right|^{-1}$, with the Poynting vector. Considering the forward-propagating RCP mode [cf.\ $k^+_{\rm R}=k_0\left(\bar{n}-\alpha\right)$ in Eq.~\eqref{Ansatz}], this condition leads to 
\begin{equation*}
    u_g>0\Rightarrow\frac{{\rm d}}{{\rm d}\omega}\left[\bar{n}(\omega)-\alpha(\omega)\right] > -\frac{\bar{n}(\omega)-\alpha(\omega)}{\omega}\,.
\end{equation*}
The above inequality, together with the dispersive models of App.~\ref{App:Lorentzian Material Dispersion}, can be used to estimate bounds on bandwidth.

Using the same parameters as in Ref.~\cite{Doskolovich2014}, Fig.\ \ref{Fig_Bandwidth} plots the parabolic fitting metric $M$, defined in Eq.\ \eqref{M Parameter} of App.\ ~\ref{App:Parabolic‑Fitting Parameter M}, as a function of normalized transverse wavenumber $k_x/k_0$. Evidently, the chirality-driven structure maintains a more accurate parabolic response over a broader angular range. Indeed, the normalized spectral bandwidth over which $M < -1$ extends to 0.010 within $k_x/k_0 \in [-0.6, 0.6]$ for the chirality-driven design, compared to only 0.004 within $k_x/k_0 \in [-0.2, 0.2]$ for the Bragg-based device---over a factor of two improvement in bandwidth, sustained across a threefold larger angular span.

This distinction arises from the different physical mechanisms underpinning the two designs. In Bragg gratings, the reflection minimum emerges from constructive interference in a periodic index profile, and thus obeys a resonance condition. This introduces an inherent trade-off: increasing the number of grating layers deepens the reflection minimum but narrows the bandwidth over which a clean quadratic response is preserved. By contrast, the chirality-based structure does not rely on a periodic modulation or resonance. Its response is engineered through parameter tuning in a uniform, non-periodic slab with a controlled dispersive response. This avoids the bandwidth-versus-reflectivity trade-off of conventional Bragg gratings and enables a broader spectral range without compromising the quality of the Laplacian-like response---even in the presence of realistic material dispersion.

\section{Meta-media implementation}
\label{Sec:Meta-Media Implementation}

At the core of the proposed mechanism lies a Bragg-like resonance in the chirality domain, governed by stopbands that emerge from parameter matching rather than conventional wavelength matching. As a result, the spectral range over which unity reflectance is achieved is centered around the design wavelength set by the slab length, but it is not governed by a wavelength/frequency domain resonance condition and is therefore not limited by the narrow spectral constraints typical of resonant systems. Although intrinsic material dispersion in the refractive index and chirality, $\bar{n}(\lambda)$ and $\alpha(\lambda)$, limits the bandwidth over which polarization-selective reflection can occur, these effects can be mitigated through appropriate material design. For example, the archetypal metamaterial of Ref.~\cite{Pendry1996}, consisting of a host medium embedded with uniformly distributed metallic Mie resonators, was shown in Ref.~\cite{Koufidis2023a} to support robust stopbands even at low free-electron densities. Although moderate spectral shifts occur, fine-tuning the electron density stabilizes these stopbands \emph{near} $\alpha = \pm \bar{n}$.

Moreover, recent advances in nanofabrication have enabled remarkably large chiral responses across a broad range of the electromagnetic spectrum. At optical frequencies, for instance, Ref.\ \cite{Zhu2018} demonstrated $\alpha \approx 0.15$ at $540~\mathrm{nm}$ using gammadion geometries. Dielectric metasurfaces have pushed these values significantly higher: optimized vertically offset dielectric bar pairs achieved $\alpha \approx 3$ over $596$--$604~\mathrm{nm}$ \cite{Gorkunov2020}, while all-dielectric toroidal dipole architectures demonstrated $\alpha \approx 4.8$ between $975$ and $995~\mathrm{nm}$ \cite{Wu2023}. At longer wavelengths, Ref.\ \cite{Zhang2009} reported $\alpha \approx 2.45$ at $0.27~\mathrm{mm}$ in a three-dimensional gold-based chiral meta-medium, and in the centimeter regime, Ref.\ \cite{Plum2009} observed $\alpha$ values between $-0.62$ and $-2.65$ at $5.08$--$6.52~\mathrm{cm}$ using a four-layer rosette-based metamaterial.

A key challenge in designing a meta-medium capable of supporting such a spectral response lies in the distinct dispersive behaviour of permittivity and chirality.  Indeed, permittivity typically exhibits Lorentzian dispersion shaped by electronic or molecular resonances, whereas chirality—originating from magnetoelectric coupling—is sharply peaked near resonance and decays rapidly off-resonance [cf.\ Eqs.\ \eqref{L_Perm_x} and \eqref{L_Chir} in App.\ \ref{App:Lorentzian Material Dispersion}]. This difference complicates simultaneous engineering of both parameters over a common frequency band. Structural approaches, however, can mitigate this disparity. Indeed, arrays of interconnected helices support continuous current pathways, hence broadening Lorentz resonances and enabling broadband optical activity with reduced dispersion \cite{Song2017}. Bilayer metasurfaces pairing chiral elements with complementary structures yield spectrally uniform optical rotation \cite{Hannam2014}, whilst recently developed planar metasurfaces sustain broadband chirality \cite{Deng2024}.

Dynamic control of chirality has only recently become experimentally viable, driven by progress in advanced 3D fabrication. In the near-infrared, triple-helical platinum nanowire metamaterials achieved tunable chirality in the range $\alpha \in (0.024, 0.032)$ across $750$--$1000~\mathrm{nm}$ \cite{Esposito2015}. At longer wavelengths, electromechanical pneumatic force produced tunability over $\alpha \in (-1.59, 0.43)$ across $0.25$--$0.37~\mathrm{mm}$ \cite{Kan2015}, and piezoelectrically actuated kirigami modulators achieved $\alpha \in (0, 1.71)$ from $0.37$ to $1.5~\mathrm{mm}$ \cite{Choi2019}. Electric-field-controlled metasurfaces reached $\alpha \in (-0.37, 0.67)$ over $1.52$--$1.75~\mathrm{\mu m}$ \cite{Zhang2021}, while conductivity-controlled designs demonstrated tunability across $\alpha \in (-0.82, 2.89)$ from $0.4$ to $5~\mathrm{mm}$ \cite{Liu2021}.

Nonreciprocal bi-anisotropic media offer further opportunities for dynamic control. Recent advances in nonlocal nonlinear metasurfaces \cite{Kolkowski2023} and space-time-modulated media \cite{Prudencio2023} suggest that substantial enhancement of nonreciprocal responses is now within reach. Moreover, certain Tellegen media---characterized by purely real magnetoelectric coupling several orders of magnitude stronger than those found in Nature \cite{Yang2025}---can reproduce key features of bandgap formation, thus enabling tunable nonreciprocal modulation through, for example, polarization currents.

Metamaterial architectures based on rosette, cross-wire, split-ring, and spiral geometries routinely achieve fractional bandwidths that exceed those of cholesteric liquid crystals (typically around $12\%$), with reported values surpassing $30\%$ (consult Table~1 in Ref.\ \cite{Koufidis2023c} and references therein). This degree of spectral tunability underscores the potential of meta-media to deliver an ``optical advantage,'' enabling passive, energy-efficient operation with extensive spatial and spectral parallelism \cite{Li2024a}. Such capabilities are particularly relevant given the increasing spatial complexity of optical computing, where recent scaling laws relate the physical dimensions of photonic systems to the computational tasks they perform \cite{Li2024b}.

\section{Conclusions}
\label{Sec:Conclusions}

This work introduces a broadband, polarization-sensitive all-optical Laplacian operator in modulus, realized via engineered ``spectral holes'' in cascaded, doubly birefringent uniform slabs with slightly mismatched refractive indices. Through coupled-wave theory analyses we have demonstrated that tuning of material parameters enables the formation of high-quality spectral holes in the chirality domain, without requiring spatial periodicity or resonance, both of which commonly limit conventional designs. Indeed, by contrast to traditional Bragg-based scattering mechanisms, where periodicity constrains bandwidth and imposes stringent fabrication tolerances, the different physical mechanism of our structure enables operation without spatial periodicity or resonant filtering, thereby relaxing fabrication constraints. Moreover, chirality introduces the additional degree of freedom necessary to enable polarization-selective Laplacian differentiation in modulus---a functionality that, to the best of our knowledge, has remained largely unexplored.

The proposed platform leverages recent advances in meta-optics, particularly in bi-anisotropic metamaterials capable of supporting giant and tunable chirality over broad spectral ranges. These developments not only enable practical realizations of the proposed edge-detection scheme but also open new directions for reconfigurable contrast enhancement and pattern recognition.

Together, these results outline a promising route towards compact, broadband, and reconfigurable platforms for all-optical analog computing. As capabilities in nanofabrication, chirality control, or even synthetic bi-anisotropy continue to mature, such metamaterials may enable not only the edge-detection device introduced here, but also multifunctional photonic components (e.g., modulators and ultra-sensitive sensors) thereby broadening the scope of integrated optics.

\section*{Funding}
S. F. K. and M. W. M. acknowledge support from a Kickstarter Research Project funded by the Department of Physics at Imperial College London. S. F. K. also gratefully acknowledges the Bodossaki Foundation for its continued financial support since 2021. Z. H. and J. B. P. were supported by the Engineering and Physical Sciences Research Council (EPSRC) under grant EP/Y015673/1. F. M. acknowledges support from the Air Force Office of Scientific Research under grant FA9550-22-1-0204, managed by Dr. Arje Nachman.

\section*{Author Contributions}
S. F. K. and Z. H. contributed equally to this work. All authors have accepted responsibility for the entire content of this manuscript and approved
its submission.

\section*{Conflict of Interest}
Authors state no conflict of interest.

\section*{Data Availability Statement}
The datasets generated during and/or analyzed during
the current study are available from the corresponding
authors on reasonable request.

\section*{Appendices}
\appendix 

\section{Cascaded Circularly and Linearly Birefringent Media}
\label{App:Cascaded Circularly and Linearly Birefringent Media}

Let $\mathbf{E}$ and $\mathbf{B}$ denote the phasors of the fundamental electric and magnetic fields, respectively, and let $\mathbf{D}$ and $\mathbf{H}$ represent their corresponding induced excitations. We may then define the auxiliary fields $\mathbf{b} = c_0 \mathbf{B}$, $\mathbf{d} = \epsilon_0^{-1} \mathbf{D}$, and $\mathbf{h} = \eta_0 \mathbf{H}$, all possessing the same units as the electric field. Here $\epsilon_0$, $\mu_0$, $\eta_0 = \left(\mu_0/\epsilon_0\right)^{1/2}$, and $c_0 = \left(\epsilon_0 \mu_0\right)^{-1/2}$ denote the vacuum permittivity, permeability, impedance, and phase velocity of light, respectively.

For a reciprocal bi-isotropic medium, Tellegen's constitutive relations in the frequency domain read 
\begin{equation*}
    \begin{pmatrix}
        \mathbf{d} \\
        \mathbf{b}
    \end{pmatrix}= 
    \begin{pmatrix}
        \epsilon_{\mathrm{r}} & i\alpha \\
        -i\alpha & \mu_{\mathrm{r}}
    \end{pmatrix}\begin{pmatrix}
        \mathbf{E} \\
        \mathbf{h}
    \end{pmatrix}\,,
\end{equation*}
with symbols as defined in the main text. For monochromatic excitation with an implicit $e^{-i\omega t}$ convention, Maxwell's macroscopic source-free curl relations become
\begin{subequations}
\begin{align}
    \nabla \times \mathbf{E} &= k_0 \alpha \mathbf{E} + i k_0 \mu_{\mathrm{r}} \mathbf{h}\,, \label{Curl Relation 1}
    \\
    \nabla \times \mathbf{h} &= -i k_0 \epsilon_{\mathrm{r}} \mathbf{E} + k_0 \alpha \mathbf{h}\,. \label{Curl Relation 2}
\end{align}
\end{subequations}

Assuming plane-wave propagation along the $z$-axis, the Bohren decomposition \cite{Lindell1994book} decouples the system of Eqs.\ \eqref{Curl Relation 1} and \eqref{Curl Relation 2} into two subsystems with orthogonal circular eigenstates. For weak chirality, $|\alpha| < {\rm Re}(n)$, the forward-propagating RCP and LCP eigenmodes are
\begin{subequations}
    \begin{equation*}\label{CB_eigen_1}
        \frac{A_{\rm R}^{+}}{\sqrt{2}} e^{i k_0(n-\alpha)z} \left(\begin{matrix}1\\-i\\\end{matrix}\right) \quad \text{and} \quad \frac{A_{\rm L}^{+}}{\sqrt{2}} e^{i k_0(n+\alpha)z} \left(\begin{matrix}1\\i\\\end{matrix}\right)\,,
    \end{equation*}
respectively, whilst the backward-propagating modes are
\begin{equation*}\label{CB_eigen_2}
    \frac{A_{\rm R}^{-}}{\sqrt{2}} e^{-i k_0(n+\alpha)z} \left(\begin{matrix}1\\i\\\end{matrix}\right) \quad \text{and} \quad \frac{A_{\rm L}^{-}}{\sqrt{2}} e^{-i k_0(n-\alpha)z} \left(\begin{matrix}1\\-i\\\end{matrix}\right)\,,
\end{equation*}
\end{subequations}
with $A_{\rm R}^{\pm}$, $A_{\rm L}^{\pm}$ being constant amplitudes. For giant-chirality, $|\alpha| \geq \mathrm{Re}(n)$, the handedness and direction of phase propagation of counter-propagating modes are interchanged, as two of the modes refract negatively \cite{Pendry2004}. 

Linear birefringence is subsequently introduced by a relative permittivity tensor. In suitable coordinates chosen to diagonalize the latter, its transverse components read $\boldsymbol{\epsilon}_\bot = \operatorname{diag}(\epsilon_1, \epsilon_2)$. Replacing the scalar permittivity $\epsilon_{\mathrm{r}}$ with the $\epst_\bot$ tensor in Eqs.\ \eqref{Curl Relation 1} and \eqref{Curl Relation 2}, the chiral Helmholtz equation for the transverse electric field reads
\begin{equation} \label{Helmholtz Wave Equation Double Chiral}
    \frac{{\rm d}^2 \mathbf{E}_\perp}{{\rm d}z^2} + 2 \alpha k_0 \frac{{\rm d}}{{\rm d}z}\left(\hat{\mathbf{z}} \times \mathbf{E}_\perp\right) + k_0^2 \left(\epst_\bot \mu_{\mathrm{r}} - \alpha^2\right) \mathbf{E}_\perp = \mathbf{0}\,,
\end{equation}
describing axial propagation in a CLB medium.

A coupled-wave model can be developed by adopting an ansatz that treats linear birefringence as a perturbation to the purely circularly birefringent eigenmodes, viz., 
\begin{align}\label{Ansatz}
    \mathbf{E}_\perp &= \left[\frac{A_L^+}{\sqrt{2}} e^{ik_0(\bar{n}+\alpha)z} + \frac{A_{\mathrm{R}}^-}{\sqrt{2}} e^{-ik_0(\bar{n}-\alpha)z}\right]\left(\begin{matrix}1\\i\\\end{matrix}\right) + \left[\frac{A_L^-}{\sqrt{2}} e^{-ik_0(\bar{n}+\alpha)z} + \frac{A_{\mathrm{R}}^+}{\sqrt{2}} e^{ik_0(\bar{n}-\alpha)z}\right] \left(\begin{matrix}1\\-i\\\end{matrix}\right)\,.
\end{align}

Substituting Eq.\ \eqref{Ansatz} into Eq.\ \eqref{Helmholtz Wave Equation Double Chiral}, the slowly varying envelope approximation leads to two sets of coupled-wave equations. Considering RCP light excitation we write \cite{Koufidis2023a}
\begin{equation}\label{CWEs}
    \frac{{\rm d}}{{\rm d}z}
    \begin{pmatrix}
        A_{\mathrm{R}}^+ \\
        A_{\mathrm{R}}^-
    \end{pmatrix}
    =
    \begin{pmatrix}
        0 & i\kappa e^{-i\delta_{\mathrm{R}}z} \\
        -i\kappa e^{i\delta_{\mathrm{R}}z} & 0
    \end{pmatrix}
    \begin{pmatrix}
        A_{\mathrm{R}}^+ \\
        A_{\mathrm{R}}^-
    \end{pmatrix}\,.
\end{equation}
For the $j$-th slab of a CLB medium, Eq.\ \eqref{CWEs} permits analytic solutions yielding a transfer matrix 
\begin{equation*}
    \mathbf{T}_{\mathrm{R}}^{(j)} =
    \begin{pmatrix}
        e^{- i\delta_{\mathrm{R}}^{(j)}{L_{j}^{-}}/{2}} p_{\mathrm{R}}^{+,(j)} & e^{- i\delta_{\mathrm{R}}^{(j)}{L_{j}^{+}/{2}}} q_{\mathrm{R}}^{+,(j)} \\
        e^{ i\delta_{\mathrm{R}}^{(j)}{L_{j}^{+}}/{2}} q_{\mathrm{R}}^{-,(j)} & e^{ i\delta_{\mathrm{R}}^{(j)}{L_{j}^{-}}/{2}} p_{\mathrm{R}}^{-,(j)}
    \end{pmatrix}\,,
\end{equation*}
where the definitions of $p_{\mathrm{R}}^{\pm,(j)}$ and $q_{\mathrm{R}}^{\pm,(j)}$ are given by Eqs.\ \eqref{CWT_P} and \eqref{CWT_Q} in the main text.

For the cascaded geometry of two equal-length slabs of CLB, the total transfer matrix $\mathbf{T}_{\mathrm{R}} = \mathbf{T}_{\mathrm{R}}^{(2)} \mathbf{T}_{\mathrm{R}}^{(1)}$, whose representative form is given in Eq.~\eqref{Total_T_Matrix}, has components:
\begin{align*}
    T_{11} &= e^{-i\delta_{\mathrm{R}}^{(2)} L/4} p_{\mathrm{R}}^{+,(2)} e^{-i\delta_{\mathrm{R}}^{(1)} L/4} p_{\mathrm{R}}^{+,(1)} +\, e^{-i3\delta_{\mathrm{R}}^{(2)} L/4} q_{\mathrm{R}}^{+,(2)} e^{i\delta_{\mathrm{R}}^{(1)} L/4} q_{\mathrm{R}}^{-,(1)}\,, 
        \\
        T_{12} &= e^{-i\delta_{\mathrm{R}}^{(2)} L/4} p_{\mathrm{R}}^{+,(2)} e^{-i\delta_{\mathrm{R}}^{(1)} L/4} q_{\mathrm{R}}^{+,(1)} +\, e^{-i3\delta_{\mathrm{R}}^{(2)} L/4} q_{\mathrm{R}}^{+,(2)} e^{i\delta_{\mathrm{R}}^{(1)} L/4} p_{\mathrm{R}}^{-,(1)}\,,
           \\
        T_{21} &= e^{i3\delta_{\mathrm{R}}^{(2)} L/4} q_{\mathrm{R}}^{-,(2)} e^{-i\delta_{\mathrm{R}}^{(1)} L/4} p_{\mathrm{R}}^{+,(1)} +\, e^{i\delta_{\mathrm{R}}^{(2)} L/4} p_{\mathrm{R}}^{-,(2)} e^{i\delta_{\mathrm{R}}^{(1)} L/4} q_{\mathrm{R}}^{-,(1)}\,,
           \\
               T_{22} &= e^{i3\delta_{\mathrm{R}}^{(2)} L/4} q_{\mathrm{R}}^{-,(2)} e^{-i\delta_{\mathrm{R}}^{(1)} L/4} q_{\mathrm{R}}^{+,(1)} +\, e^{i\delta_{\mathrm{R}}^{(2)} L/4} p_{\mathrm{R}}^{-,(2)} e^{i\delta_{\mathrm{R}}^{(1)} L/4} p_{\mathrm{R}}^{-,(1)}\,. 
\end{align*}

\section{Parabolic‑Fitting Parameter $M$}
\label{App:Parabolic‑Fitting Parameter M}
Here we quantify how closely the transfer‑function magnitude $\lvert R(k_x)\rvert$
follows an ideal quadratic $a\,k_x^2$ over the range $|k_x|\le k_{x,\max}$; $a$ is a free parameter. We begin by defining the residual variance
\begin{equation*}
S_{\mathrm{res}}
=\int_{-\,k_{x,\max}}^{+\,k_{x,\max}}
\left[\lvert R(k_x)\rvert - a\,k_x^2\right]^{2}\,\mathrm{d}k_x\,,
\end{equation*}
and the total variance
\begin{equation*}
S_{\mathrm{tot}}
=\int_{-\,k_{x,\max}}^{+\,k_{x,\max}}
\left[\lvert R(k_x)\rvert - \overline{\lvert R\rvert}\right]^{2}\,\mathrm{d}k_x\,,
\end{equation*}
where
\begin{equation*}
   \overline{\lvert R\rvert}
=\frac{1}{2\,k_{x,\max}}
\int_{-\,k_{x,\max}}^{+\,k_{x,\max}}
\lvert R(k_x)\rvert\,\mathrm{d}k_x 
\end{equation*}
is the mean magnitude in that interval.  The coefficient of determination is then
\begin{equation*}
\rho^2
=1 - \frac{S_{\mathrm{res}}}{S_{\mathrm{tot}}}\,,
\end{equation*}
so that the parabolic‑fitting parameter $M$ is defined by
\begin{equation}\label{M Parameter}
M 
=\log_{10}\bigl(1 - \rho^2\bigr)
=\log_{10}\!\Bigl(\frac{S_{\mathrm{res}}}{S_{\mathrm{tot}}}\Bigr)\,.
\end{equation}

\noindent A more negative value of $M$ indicates a closer match to the ideal
quadratic dependence.  The choice of $k_{x,\max}$ determines the maximum
normalized transverse wavenumber over which the fit is performed.

\section{Lorentzian Material Dispersion}
\label{App:Lorentzian Material Dispersion}

The broadband response in Fig. 5 is computed using a more realistic material model in which both the permittivity tensor and the chirality parameter follow a common Lorentzian dispersion, namely
\begin{subequations}
\begin{align}
    \epsilon_{xx}(\omega) &= \epsilon_\infty + \frac{F_\epsilon}{\omega_0^2 - \omega^2 - i \Gamma \omega}\,, \label{L_Perm_x}
    \\
    \epsilon_{yy}(\omega) &= \epsilon_\infty + \frac{F'_\epsilon}{\omega_0^2 - \omega^2 - i \Gamma \omega}\,, \label{L_Perm_y}
    \\
    \alpha(\omega) &= \frac{F_\alpha \, \omega (\omega_0^2 - \omega^2)}{(\omega_0^2 - \omega^2)^2 + (\Gamma \omega)^2}\,. \label{L_Chir}
\end{align}
\end{subequations}
Here $\omega_0 = 2\pi c_0 / \lambda_{\mathrm{0,r}}$ is the resonance frequency, $\Gamma = 0$ is the damping rate, and $\epsilon_\infty=1$ is the high-frequency limit. In Fig.~5 the resonance wavelength is chosen as $\lambda_{\mathrm{0,r}} = 2\lambda_\mathrm{0}$.

At the design wavelength $\lambda_\mathrm{0}$, the real parts of the permittivities are constrained to
\begin{align*}
\text{Slab 1:} &\quad \text{Re}\{\epsilon_{xx}\} = 4.0\,, \\
\text{Slab 2:} &\quad \text{Re}\{\epsilon_{xx}\} = 6.0\,, \\
\text{Both slabs:} &\quad \text{Re}\{\epsilon_{yy}\} = 4.95\,,
\end{align*}
thereby determining the oscillator strengths via
\begin{equation*}
F_\epsilon = (\epsilon_\mathrm{target} - \epsilon_\infty) \cdot \frac{|\omega_0^2 - \omega_d^2 - i \Gamma \omega_d|^2}{\operatorname{Re}(\omega_0^2 - \omega_d^2 - i \Gamma \omega_d)}\,,
\end{equation*}
where $\omega_d = 2\pi c_0 / \lambda_\mathrm{0}$. The chirality strength $F_\alpha$ is computed in the same way from the target value $\alpha(\omega_d) = \alpha_c = 2.228$ that satisfies the spectral-tuning condition.

This dispersion model ensures that the target values for the medium parameters are met exactly at $\lambda_0$, while remaining physically causal and passive away from resonance.

\bibliography{sorsamp.bib}

\end{document}